\documentclass[conference]{IEEEtran}
\IEEEoverridecommandlockouts

\usepackage{cite}
\usepackage{amsmath,amssymb,amsfonts}
\usepackage{algorithmic}
\usepackage{graphicx}
\usepackage{textcomp}
\usepackage{xcolor}

\usepackage[inline]{enumitem}
\usepackage{subcaption}
\usepackage{booktabs}
\usepackage{threeparttable}
\usepackage{siunitx}
\usepackage{url}
\usepackage{xurl}
\usepackage[colorlinks=true,linkcolor=blue,citecolor=blue,urlcolor=blue]{hyperref}

\usepackage{xspace}
\newcommand{\ie}{{\em i.e.,}\xspace}
\newcommand{\eg}{{\em e.g.,}\xspace}
\newcommand{\etal}{{\em \ et al.}\xspace}

\def\BibTeX{{\rm B\kern-.05em{\sc i\kern-.025em b}\kern-.08em
    T\kern-.1667em\lower.7ex\hbox{E}\kern-.125emX}}
\begin{document}

\title{Robustness Analysis of Australia's Internet Using a Multilayer Network Model}

\author{\IEEEauthorblockN{Benjamin Lang \qquad Matthew Roughan \qquad  Mengbin Ye}
\IEEEauthorblockA{\textit{Adelaide Data Science Centre, School of Mathematical Sciences, Adelaide University, Australia}\\
\{benjamin.lang, matthew.roughan, ben.ye\}@adelaide.edu.au}
}

\maketitle

% ABSTRACT
\begin{abstract}
Australia depends on an Internet built from multiple networks of long-haul links. We study the interactions of these independent provider networks to investigate how the peering between these networks provides redundancy for failures on a single network, as well as the potential vulnerabilities introduced by failures of Shared Risk Link Groups (SRLGs), whereby ostensibly independent links of different providers fail simultaneously due to joint physical dependencies such as shared conduits.

We introduce a generalised multilayer network model in which each layer represents the network of an individual Internet Service Provider (ISP), along with an Internet Exchange Point (IXP) layer that facilitates interconnections between ISP networks. We construct an Australia-specific model, consisting of six major ISPs. A failure analysis is performed on this network, revealing that diversity provides redundancy, even in the presence of shared risks, indicating the importance of a diverse network ecosystem.
\end{abstract}


%%% INTRODUCTION %%%
\section{Introduction}
\label{sec:intro}
The Internet is a critical infrastructure of the modern world. It is often taken for granted that current infrastructure will operate with continued performance and reliability. But disruptions to the network are not uncommon, with over 180 occurring globally in 2025 \cite{Belson_2026}. The causes of failures vary from intentional acts (\eg terrorism, government shutdowns and cyber-attacks), to accidental fibre cuts, natural disasters and power failures \cite{Belson_2026, Baumann_2014, Manzano_2012, Rueda_2017, Durairajan_2015}. Multiple failures on the network can lead to major outages, affecting millions of users and services worldwide \cite{Qiu_2011}. It is therefore increasingly important to understand how vulnerable the Internet is to such failures. 

Progress to understand the Internet's structure and vulnerabilities has been hindered by its complexity, scale, and diversity of ownership. Mapping is constrained by the limitations of existing measurement tools~\cite{Augustin_2006, Roughan_2011, Willinger_2013, Motamedi_2015} and by Internet Service Providers (ISPs) restricting access to network data due to security concerns. Previous studies have collected publicly available high-level maps from network providers, such as the Topology Zoo \cite{Knight_2011} and Internet Atlas \cite{Durairajan_2013}. However, these topologies do not account for all of the interdependencies between provider networks. 

\begin{figure}[!t]
    \centering
    \includegraphics[width=0.29\textwidth]{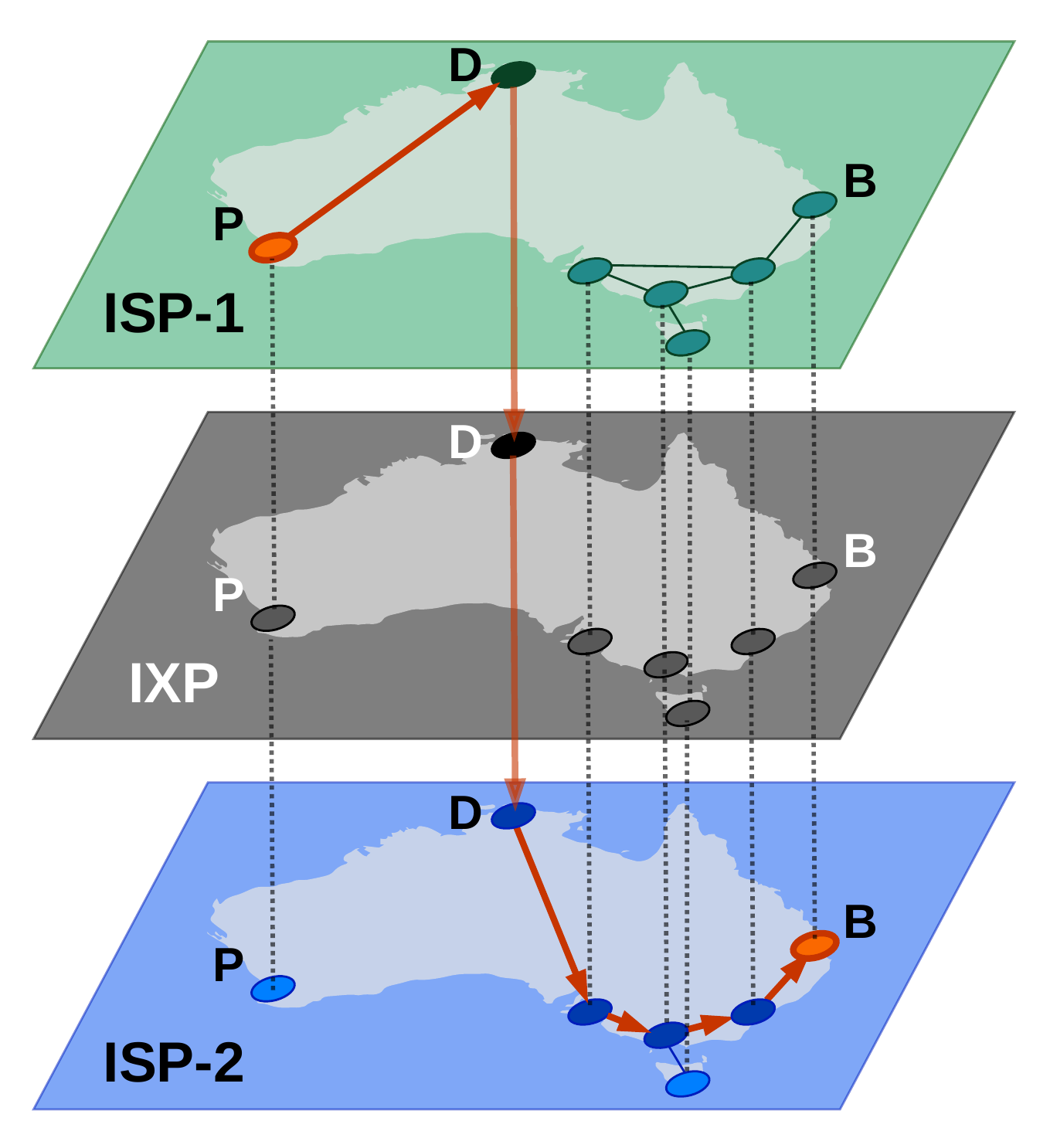}
    \caption{Simplified view of the multilayer framework for Australia. A viable path between layers is shown, providing a route from Perth (Node P) to Brisbane (Node B), despite this route not being available through ISP-1 or ISP-2 alone.}
    \vspace*{-4mm}
    \label{fig:multi-model}
\end{figure}

We investigate two specific dependencies, illustrated in \autoref{fig:multi-model}. Firstly, Internet eXchange Points (IXPs) are centralised physical locations for providers to connect and exchange traffic~\cite{Roughan_2022}. IXPs change a network's robustness by providing hubs for interconnections in major cities. This enables connections between parallel provider networks, increasing the number of paths between locations. Secondly, Shared Risk Link Groups (SRLGs) are a set of network links that share physical infrastructure. Examples of SRLGs include two links being housed in the same conduit, or sharing wavelengths on the same fibre \cite{Shao_2010}. A single SRLG failure can cause all links within the group to simultaneously fail. Therefore, ostensibly independent links of different provider networks may appear to provide redundancy, but in fact obscure serious vulnerabilities due to shared failures~\cite{Carter_2002}.

This motivates our central question: \textbf{How robust is the Internet to failures of SRLGs, and do diverse ISP networks provide redundancy thanks to the presence of IXPs?}

Specifically, we are interested in national robustness so we concentrate on wide-area networks and on long-haul fibre-optic cables rather than wireless communications. 

This paper provides three key contributions: 
\begin{itemize}

    \item We propose a multilayer network framework to model the multiple networks of a Region of Interest (RoI). The multilayer network uses two types of layers: ISP layers, with each layer comprising an individual provider's independent network from the RoI; and a single IXP layer, used to capture peering behaviour between ISPs. 

    \item We demonstrate this multilayer network by applying it to Australia. This forms a good case study: the scale is such that a nationwide analysis is tractable but still of sufficient complexity to be interesting. Australia's geographical isolation, vast distances, and population distribution result in an interesting variety of topologies. Additionally, the Internet landscape contains a mix of public and private ownership, dominated by a relatively small number of retail ISPs, but with enough independence for interest. And though studies have been conducted for other nations, such as the United States \cite{Durairajan_2015}, there are few detailed studies of mid-sized countries, despite their prevalence.

    \item We perform a robustness analysis of the Australian network, using an exhaustive analysis of all single and double SRLG failures.
    
\end{itemize}

The results show that (1) the ecosystem of ISPs and IXPs is more robust than any individual network (avoiding 86\% of disconnects), despite the presence in our model of extensive and conservatively estimated SRLGs, and (2) there are, nevertheless, points of vulnerability--for example, connections to and within Tasmania, where all networks rely on the same low-diversity paths.

%%% RELATED WORK %%%
\section{Related Work}
\label{sec:related_work}
% Resilience, reliability, robustness definitions
Here we use the term {\em robustness} to refer to the ability of a system to continue operating under unexpected disruptions~\cite{Beyza_2022}, aligning with many failure analyses, \eg \cite{Durairajan_2015, Faber_2017, Maniadakis_2013, Manzano_2012}, rather than the terms reliability \cite{Vaisman_2021, Neumayer_2010} and resilience \cite{Hall_2013, Faber_2017}. For our analysis, we align this with a graph-theoretic definition, whereby a network is robust to link removals if the network remains connected (a path exists between any chosen node pair) \cite{Manzano_2012} despite the removed component.

% failure analyses
Earlier topology studies primarily focused on simple graph-theoretic approaches, examining Internet robustness using node removal \cite{Doyle_2005, Xiao_2008, Baumann_2014} and standard graph metrics to evaluate the impact of failures, such as the size of largest connected component. These strategies were applied to synthetic topologies \cite{Doyle_2005, Xiao_2008}, measured topologies \cite{Xiao_2008, Baumann_2014}, and real-world topologies \cite{Rueda_2017, Manzano_2012,Knight_2011}. We focus on link loss because our interest is with wide-area networks and SRLGs. A previous link failure study includes a resilience analysis of the global submarine cable system \cite{Omer_2009}. 

% SRLGs
There are relatively few studies considering SRLG failures, largely because it is difficult to access the underlying data. Example studies include 
\cite{Neumayer_2010,Durairajan_2015, Shao_2010, Vaisman_2021}. The InterTubes study~\cite{Durairajan_2015} offers the closest comparison to our work: a national study of the United States long-haul network, analysing  the impact of SRLG failures using real-world maps from providers. This approach inspires aspects of our failure analysis, specifically in \autoref{subsec:shared-risk-results}. However, our proposed multilayer model explicitly incorporates IXPs and peering relations between ISP networks, and we perform a robustness analysis specific to a mid-sized country, Australia. 

% national models
National or regional-based Internet models are also uncommon. Previous models have used end-to-end measurements or traceroute to infer maps, such as in China \cite{Tian_2012} and Japan \cite{Yoshida_2009}, or use existing traceroute datasets \cite{Anya_2025}. 
These studies treat ISP networks as independent, and do not account for interactions between networks \cite{Tian_2012, Yoshida_2009}. A national model of Australia's network has not yet been developed.

Studies of IXPs have primarily focused on their coverage in traceroute datasets \cite{Anya_2025} or on mapping peering interconnections to specific locations \cite{Giotsas_2015}, rather than their role within topology models. As revealed in our analysis of the Australian network, IXPs play a central role in maintaining the robustness of the overall network, even during SRLG failures.

%%% METHODS %%%
\section{Model Development}
\label{sec:method}
Here we propose a multilayer network framework that can be applied to other nations or geographical regions. To illustrate the framework, we apply our approach to Australia. The relevant data and code used in subsequent analysis is located on GitHub: \url{https://github.com/lang-b/au-multilayer-model/}.

\subsection{Multilayer Network Framework}
An example of our multilayer network is shown in \autoref{fig:multi-model}. The model uses two types of layers: 
\begin{enumerate}
    \item ISP layers, each corresponding to the topology of an ISP backbone network in the region; and 
    \item an IXP layer, depicting locations of IXPs. 
\end{enumerate}
Inter-layer links occur only between an ISP layer and the IXP layer, between nodes in the same location (\ie the same city on both layers). These inter-layer links allow for routes between different ISPs, emulating peering through IXPs\footnote{An important caveat is that we do not incorporate inter-domain routing constraints, and so there may be paths that are physically possible, but not usable. Future work should investigate the additional constraints so imposed.}. Note, each ISP layer corresponds to a single ISP network. Any number of ISP layers can be used in this model. Each such layer need not include all possible nodes. 

An example path is described in \autoref{fig:multi-model}. A path from Perth to Brisbane is not possible on either ISP layer, but inter-layer links through the IXP layer create a viable path. The figure is simplified: the networks we consider below are initially connected, and may cover all the nodes nationwide, but become unconnected under failures. 

The nodes are {\em Metropolitan Points of Presence} (MPoPs): the consolidated network equipment of an ISP across a metropolitan area\footnote{The definition of a metropolitan area is challenging, and ISPs may differ in how they represent such, for instance, by including small regional centres into a larger metro area. However, the majority of traffic is well-served by agreed definitions in Australia.}. Modelling at the MPoP level within Layer 3 is beneficial, providing geographic information, while maintaining a higher level of abstraction than modelling at the router or interface level \cite{Roughan_2011}, where data is unavailable or untrustworthy, and redundancy levels are typically higher in any case. Links then correspond to inter-city optical fibre, including both terrestrial and submarine cables. 

Although Australia has multiple organisations that provide IXPs, we use a single IXP layer with nodes in cities determined by public data  (a list is provided in the dataset). Additional IXP layers would only provide node redundancy and we consider link redundancy here. The IXP layer has no internal links---it connects other layers.

% definition of network
We use the mathematical framework of Kivel\"a\etal\cite{Kivela_2014}. Each layer is an undirected multigraph $G=(V,E)$, where $V$ is a set of nodes $V=\{v_1, v_2, \ldots , v_N \} $ and $E$ is a list of two-tuples from $V \times V$ that we refer to here as {\em links} (though the term edge is often used mathematically). We generalise this to a multilayer network captured by the tuple $M=\left(V_M,E_M, V, L \right)$, where $L=\{x, y_1, y_2, \ldots, y_d  \}$ consists of $x$, the IXP layer, and the ISP layers $y_i$, and $V$ is the union of all of the nodes present in each layer (here, the set of all MPoP locations, represented by cities and towns).  

The layer-specific components are then formed over all valid tuples of (MPoP, layer) denoted as $V_M\subseteq V\times L$. For instance $(u,y_i) \in V_M$ indicates node $u$ from $V$ exists in network layer $y_i$. Then $E_M\subseteq V_M \times V_M$ is the set of all links.
 
 For layers $y_1, \hdots, y_d$, the intra-layer links are given by $E_A= \big\{((u,\alpha),(v,\beta))\in E_M \vert \alpha = \beta \big\}.$ The network has no intra-layer links for layer $x$.
Inter-layer links correspond to links between ISP layers and the IXP layer, and are particularly rare for small to medium ISPs such as those of a mid-sized country like Australia. Although it is possible for ISPs to directly interconnect, substantial work shows that this type of interconnect is becoming relatively rare in the modern Internet~\cite{Giotsas_2015}. Thus, we assume no inter-layer links occur directly between ISP layers. 
Inter-layer links only occur at the same node $u$ in $V$ between an IXP and ISP layer,  so they form the sets $E_{x, y_i} = \big\{((u,\alpha),(v,\beta))\in E_M \mid u=v, \beta=x, \alpha \in y_i \big\}$.

Our multilayer network framework includes a set of SRLGs: $S=\{S_1, S_2, \ldots, S_m \},$ where $S_i\subseteq E_A$ is a given SRLG. 

We next demonstrate the model with an Australian case study, though 
it can be adapted to other nations or regions. 

\subsection{Australian Model}
\label{subsec:data}
We create the Australian model by \begin{enumerate*}
    \item sourcing publicly available maps from providers; 
    \item using IXP shared data to identify ISPs with a presence at each; and 
    \item assign layer links to SRLGs based on observed common transport infrastructure, such as roads, railways, or a specific submarine cable.
\end{enumerate*}

\subsubsection{Topology Collection}
ISP topologies were derived from publicly available ISP network maps. The collected maps and associated metadata are available at \url{https://github.com/lang-b/au-multilayer-model/}. The maps were manually transcribed, and metadata was recorded where available (\eg whether a link is planned or operational, or whether a link is owned by a third-party).  

The final model includes six ISP layers, consisting of their Australian network topologies. This includes five retail service providers: Telstra, Optus, Vocus, Aussie Broadband (ABB), and Superloop; and Australia's only academic network provider, AARNet. The retail providers are five of the top six providers based on wholesale market share in December 2025 \cite{ACCC_2026}\footnote{TPG has the second-highest market share in Australia, but we could not locate a corresponding network map. Due to Vocus' acquisition of TPG's fibre assets in 2025 \cite{Vocus_2025}, we argue that Vocus and TPG will use the same fibre network, at least in the future, and are therefore grouped in our analysis.}.  Our model therefore accounts for 92.5\% of Australia's Internet based on market share, in addition to Australia's academic network. The number of nodes and links for each ISP layer are provided in \autoref{tab:model-metrics}. 

Map fidelity varies significantly. For instance, AARNet's network includes 33 regional nodes, while Telstra's only shows the eight state capitals. Due to our interest in major outages, we focus on state capitals and regional towns that feature consistently across ISP maps and which were also necessary to preserve the network's topology. For example, Alice Springs featured in all six maps; however, as a transit node between two major cities, it was omitted from the model. In contrast, Launceston was retained because it connects Tasmania to the mainland. In general, we sought to preserve the essential topology (often including nodes with 3 or more links, that are central to the network), while excluding transitory or peripheral regional nodes (i.e., nodes connected to only a single other node) in order to maximise the simplicity of the resulting network.

The final list of nodes included in the model is provided in the dataset, denoted by their 3-letter airport codes. Represented are 8 capital cities and 5 regional centres. Australia has a highly skewed population distribution, and so these 13 centres still represent at least three quarters of Australia's population.  

Additionally, some maps aligned links with their routes, while others took more artistic license. The quality of metadata also varied between maps. Finally, we note that some links were indicated as {\em planned} in some form. We recorded metadata in the model, and still included planned links so that the model would not become quickly outdated. 

\begin{figure*}[!t]
    \centering
    \begin{subfigure}{0.54\textwidth}
        \centering
        \includegraphics[width=\textwidth]{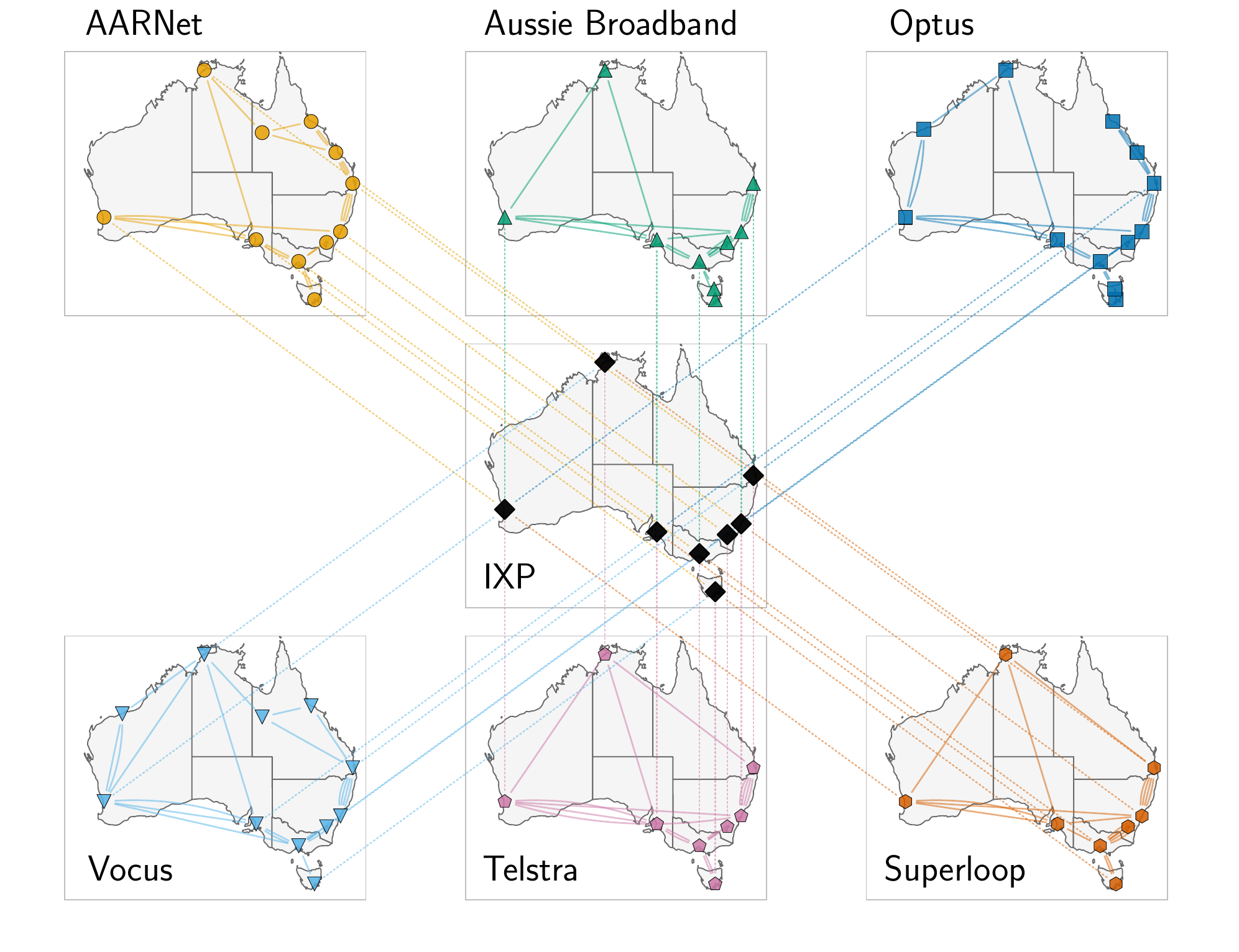}
        \caption{Australian-specific multilayer model, showing all six ISP layers and the central IXP layer. This shows the central role of the IXP layer as a hub to connect other ISP layers.}
        \label{fig:multi-model-hs}
    \end{subfigure}
    \hfill
    \begin{subfigure}{0.44\textwidth}
        \centering
        \includegraphics[width=\textwidth]{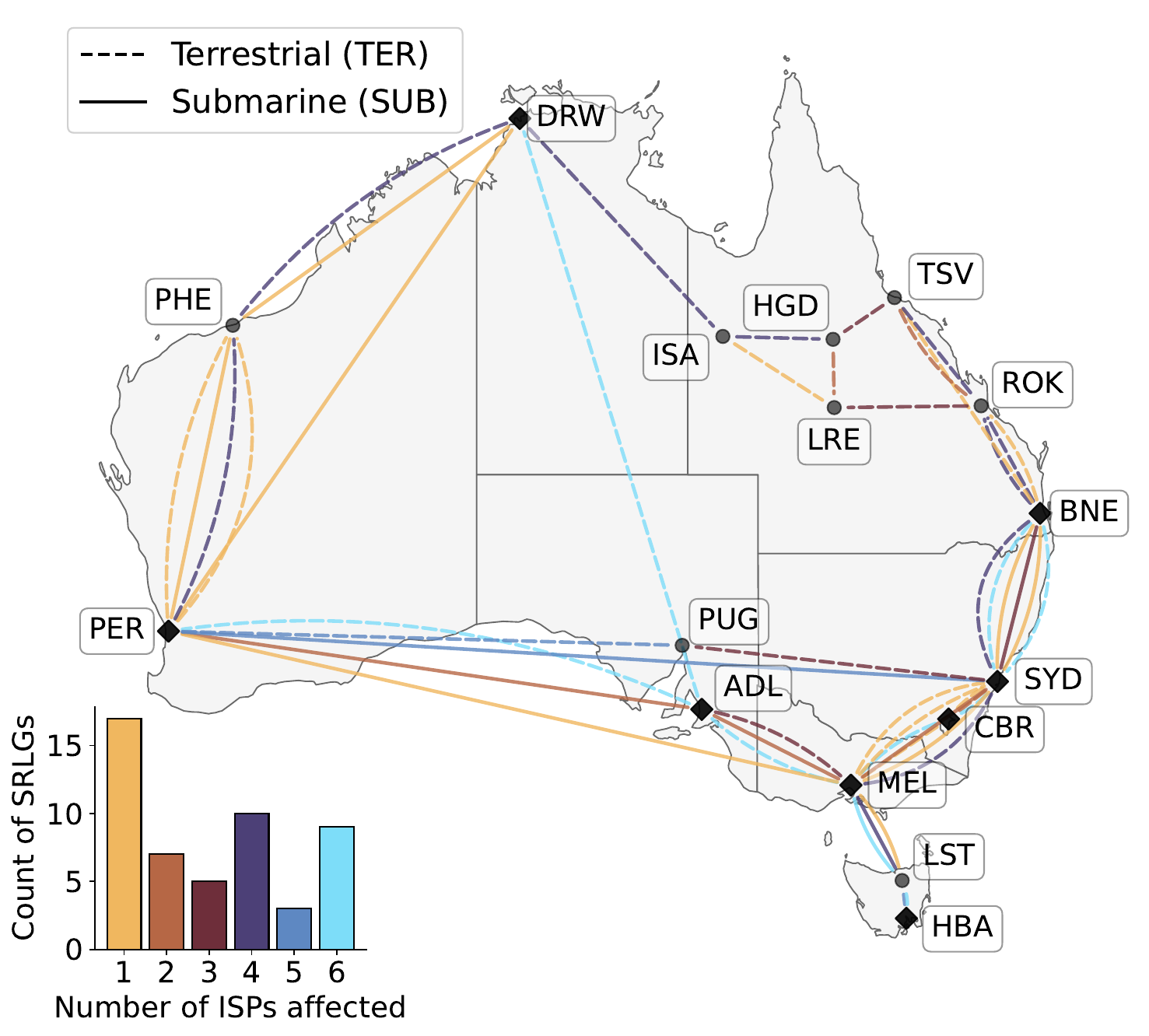}
        \caption{SRLG locations, with the colour of each SRLG denoting how many ISPs would be affected if said SRLG was removed. The inset graph denotes the number of SRLGs corresponding to each count of ISPs. Nine SRLGs are found to be used by all six ISPs, such as the link between DRW and ADL.}
        \label{fig:srg-isp-map}
    \end{subfigure}
    \caption{Final Australian multilayer model and corresponding map of SRLGs.}
    \label{fig:final-model}
\end{figure*}

\begin{table*}[!t]
\caption{Metrics for each of the model ISP layers, IXP layer, and full network.}
\label{tab:model-metrics}
\centering
\begin{tabular}{l|S[table-format=2.2] S[table-format=2.2] S[table-format=2.2] S[table-format=2.2] S[table-format=2.2] S[table-format=2.2] | r S[table-format=3.2]}
\toprule
{Metric} & {AARNet} & {Aussie Broadband} & {Optus} & {Superloop} & {Telstra} & {Vocus} & {IXP} & {Full Network} \\
\midrule
Number of nodes & 11 & 9 & 12 & 8 & 8 & 11 & 8 & 67 \\
Number of intra-layer links & 23 & 20 & 23 & 20 & 20 & 25 & 0 & 131 \\
Number of inter-layer links & 8 & 5 & 5 & 7 & 8 & 7 & 40 & 40 \\
Number of SRLGs & 28 & 22 & 24 & 24 & 29 & 28 & 0 & 51 \\
Mean intra-layer node degree & 4.18 & 4.44 & 3.83 & 5.00 & 5.00 & 4.55 & 0 & 3.91 \\
Network diameter & 5 & 3 & 5 & 3 & 3 & 4 & - & 8 \\
Clustering coefficient & 0.27 & 0.64 & 0.31 & 0.39 & 0.29 & 0.52 & - & 0.28 \\
\bottomrule
\end{tabular}
\vspace*{-4mm}
\end{table*}

\subsubsection{IXP Locations and Peering Relationships}
IXP locations were determined using the Internet Society's IXP Tracker \cite{IS_2026}. These were cross-checked using data from PeeringDB \cite{PeeringDB_2026}. The result was an IXP node in each capital.  

Most inter-layer links were inferred based on an ISP's presence at each IXP, as determined using PeeringDB. 
Telstra and Optus, however, do not advertise their peering agreements (official agreements to peer between providers), so we instead estimate IXP presence based on their public peering policies: general policies for an ISP that specify the minimum requirements other providers must fulfil in order to enter an agreement with them. These minimum requirements include a list of required peering locations and cities. Therefore, we discern these locations as being likely locations for Telstra and Optus to peer with other ISPs.

Finally, for simplicity, we consider all ISPs connecting to an IXP node to connect through the IXP with one another at that location. This allows us to utilise the single IXP layer, rather than linking ISP layers directly. This is reasonable for the ISPs who publicly advertise peering. For Telstra and Optus, we validated their peering with other ISPs using public resources. This includes peering between ABB and Telstra \cite{ABB_2026}, Vocus and Telstra \cite{ACCC_2018}, and Optus and Telstra \cite{ACCC_2018}. This verification is further used to cross-check results from PeeringDB, since the accuracy of PeeringDB data is contested \cite{Kloti_2016}. 

\subsection{SRLG Identification}
\label{subsec:srlg}
% How follow infrastructure
\subsubsection{Methodology}
We lack data on exact physical deployment of specific fibres and conduits, so we assign ISP links to SRLGs based on the assumption that terrestrial links are co-located with roadway and railway infrastructure. Fibre providers need rights-of-way and infrastructure to develop fibre corridors, and so it is impractical for long-haul terrestrial fibres to ``go it alone" \cite{Durairajan_2015,Yoshida_2009}. 
We assume a single fibre corridor per transport infrastructure, and create SRLGs per road/railway segment. It is possible that fibre follows either side of a road, creating limited diversity, but we conservatively consider this a single corridor. 

%Criteria
We create SRLGs by determining the routes taken by each link and looking for overlap. We make the conservative assumption that links are included in an SRLG if even a single section overlaps. The SRLG itself corresponds to that specific physical section. If an SRLG fails, the whole link is removed, rather than just the overlapping segment.

% difficult to determine exact route 
We assigned links to SRLGs manually. For terrestrial cables (TER), we identify the specific major roadways or railways the link follows based on the map's link position compared with the `Major Roads' \cite{DAA_2026} and `Railways' \cite{DAA_2024} datasets. We also use Google Maps to determine which routes are considered major roadways across large geographic distances. For submarine cables (SUB), we use the Submarine cable map~\cite{TeleGeography_2026}. An SRLG is assigned per submarine cable, which is simpler than the terrestrial case because there are relatively few (domestic) submarine cables and often the exact cable used appears in the metadata. If (as in a small number of cases) routes were unclear, we utilise supplementary open source information, such as social media posts and reports by ISPs. We also supplemented our work with additional fibre cable maps such as National Broadband Network (NBN) deployment maps, other government documents, and Indefeasible Right of Use (IRU) agreements with other key fibre providers where available. Information about the provenance of each SRLG is included in the dataset.

% \subsubsection{Additional SRLG Nodes}
While identifying SRLGs for the Australian model, we included three additional regional towns that were not included as nodes in the original model. These three towns are: Hughenden (HGD), Longreach (LRE), and Port Augusta (PUG). The locations of these towns are also shown in \autoref{fig:srg-isp-map}. These towns were necessary to further segment specific routes. For instance, numerous ISP maps included links between Rockhampton and Mount Isa; in reality, these routes likely either go directly from Longreach to Mount Isa, or pass through Hughenden. Therefore, it was necessary to add both towns, such that we can assign links to the appropriate SRLGs. 

% how use conservative or ideal
\subsubsection{Ideal and Conservative SRLG Sets}
Despite efforts to discern exact routes for each link, for some maps it is unclear whether links belong to an existing SRLG, or if they follow an independent route. To account for this, we defined two sets of SRLGs: an ideal set, with these ambiguous links belonging to their own independent SRLG; and a conservative set, where these links are still mapped onto existing SRLGs. Ultimately, the conservative set provides a worst-case failure analysis. We found the conservative and ideal results largely the same, and therefore focus on the conservative model here.

\subsection{Final Model and SRLGs}
\label{subsec:final-model}
The final Australian multilayer network model is shown in \autoref{fig:multi-model-hs}. Layer metrics such as number of nodes and links are shown in \autoref{tab:model-metrics}. Note that each layer is connected. 

\begin{figure}[!t] 
    \centering
    \includegraphics[width=0.33\textwidth]{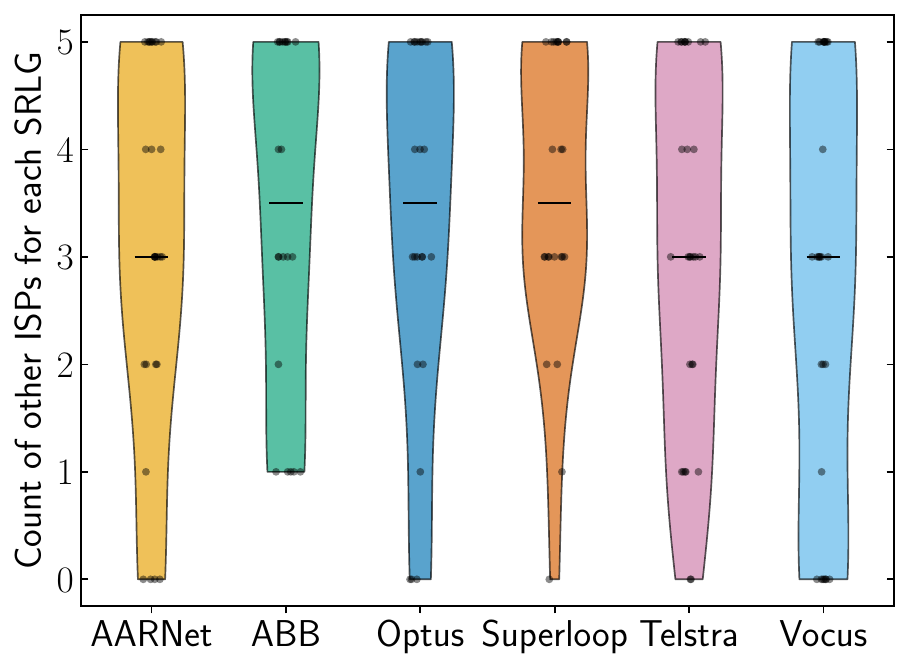}
    \caption{Distribution of SRLGs shared by each ISP: for each ISP, we count the number of other ISPs with which they share their SRLGs. Aussie Broadband and Superloop share the most SRLG infrastructure, while Vocus shares the least.}
    \label{fig:isp-shared-violin}
    \vspace*{-4mm}
\end{figure}

\autoref{fig:multi-model-hs} shows that the layers vary more than one might think. \autoref{fig:multi-model-hs} also demonstrates the central behaviour of the IXP layer as a hub for all other layers. Notice how nodes within the IXP layer do not connect with one another. We observe that inter-layer links are always present for the five most populous cities in Australia (Sydney, Melbourne, Brisbane, Perth, Adelaide). Canberra has the fewest inter-layer links of the state capitals, followed by Darwin and Hobart. 

The final map of the conservative SRLGs is shown in \autoref{fig:srg-isp-map}. The colour of each link denotes the number of ISPs that are members of an SRLG (\ie how many ISPs would be affected if said SRLG were removed). The line-style indicates if a link is terrestrial or submarine (note that the exact routes taken by the SRLG is not shown in the figure). \textbf{Of the 51 SRLGs, 20 followed roadways, 15 followed a joint railway/roadway, while 16 corresponded to submarine cables}. No route was found to follow a railway only.  

The inset bar plot in \autoref{fig:srg-isp-map} shows the counts: we observe that all six ISPs are impacted simultaneously by nine of the SRLGs (pale blue), \ie around 18\% of SRLGs cause a failure on all of the networks. These SRLGs correspond to some of the most common routes and major roads. We also note that these nine SRLGs form a path between all eight state capitals. Around a third (17) of SRLGs affect only a single ISP.   

%%% RESULTS %%%
\section{Model Analysis}
\label{sec:results}
We perform two sets of analysis on our Australian model. Firstly, we analyse the SRLG commonalities across the ISPs; then, a failure analysis by considering the removal of SRLGs for both single and double failures. 

\begin{figure}[!t]
    \centering
    \includegraphics[width=0.30\textwidth]{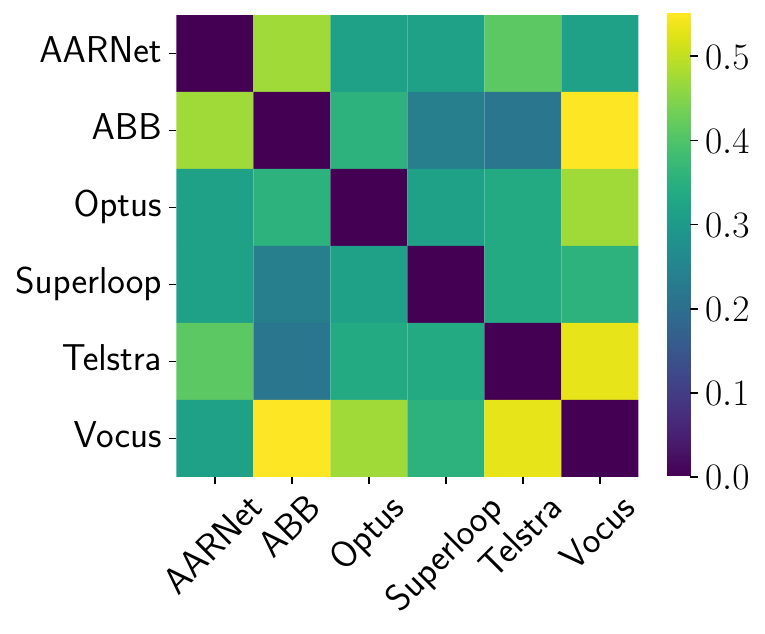}
    \caption{Dissimilarity profiles of SRLGs between ISPs, calculated using Hamming Distance. A higher value indicates ISP networks are more dissimilar. Aussie Broadband and Telstra are the most similar to one another, while Vocus is the most dissimilar from other ISP networks.}
    \label{fig:isp-heatmap}
    \vspace*{-4mm}
\end{figure}

\subsection{Shared Risk Analysis}
\label{subsec:shared-risk-results}
We analyse robustness using a similar approach to Durairajan\etal\cite{Durairajan_2015} by creating a binary indicator matrix with SRLGs as rows and ISPs as columns, set to 1 if the ISP contributes a link to that SRLG. From this we generate \autoref{fig:isp-shared-violin}, which shows for each ISP and their SRLGs, the number of other providers they tend to share with. Zero corresponds to SRLGs shared with no other providers and five indicates SRLGs shared with all other providers. Overall we find the distributions are largely the same for each ISP, with median share per SRLG of 3 or 3.5 additional ISPs. The figure shows that Vocus appears to share the least SRLGs, and Superloop appears to share the most infrastructure, with Aussie Broadband sharing SRLGs with at least one other. 

\autoref{fig:isp-heatmap} shows how dissimilar each ISP's SRLG profile is. It 
shows the Hamming Distance for each row of the indicator matrix. 
The figure shows Telstra and Aussie Broadband have the most similar topologies and Vocus appears to be the most independent. These dissimilarities are important because they underpin the redundancy observed in the failure analysis. 

\subsection{Failure Analysis}
\label{subsec:failure_results}

% \subsubsection{Methodology}
\label{subsec:failure_method}
We assess the robustness of the multilayer network model by considering the impact of failed SRLGs on the connectivity of the network. 
We perform two analyses: single and double (paired) SRLG failures. SRLG failures are non-trivial events. Double failures have precedent, \eg \cite{Belson_2026, Hales_2022}, but are rare. Triple+ failures fall outside of the scope of any typical analysis. 

We exhaustively explore all single and double SRLG failures. In each case, we remove the SRLG links, and determine the network connectivity. Note a link may belong to more than one SRLG, and it will fail in each group \cite{Vaisman_2021}.
 
We distinguish two types of disconnects: {\em local}, where an ISP layer is disconnected within its intra-layer network, but due to the inter-layer links the multilayer network remains connected (\ie all nodes still have reachable paths); and {\em global}, where the whole multilayer network becomes disconnected.  

In the event of a single SRLG failure, we find that \textbf{out of all 51 SRLGs, only two local disconnects occurred (3.9\%)}, corresponding to a Melbourne -- Launceston and Launceston -- Hobart segment. However, \textbf{no global disconnects occur due to a single SRLG failure}. This is entirely expected: most Australian ISPs plan enough redundancy to cope with most single failures, and the parallel networks easily provide sufficient redundancy in other events.

The results of double SRLG failures are shown in \autoref{tab:double-fails}. There are 1275 pairs: no disconnects occur for 1164 of these pairs (91.3\% of pairs), and in 96 cases (7.5\%), a local disconnect occurs within at least one ISP layer, but global failure is avoided. Only 15 SRLG pairs (1.2\% of pairs) produce a global disconnect. {\bf Thus, 86\% of potential global disconnects were {\em prevented} by connectivity through the other layers.}

\autoref{tab:double-fails} also shows results for the number of SRLGs that are involved in disconnects during the double SRLG failures. Immediately above, we identified two SRLGs whose removal leads to a local disconnect: $S_{M,L}$ and $S_{L,H}$. When studying the removal of pairs of SRLGs, we exclude $S_{M,L}$ and $S_{L,H}$ when analysing local disconnects, since we are already aware of their impact.
However, we do consider these two SRLGs when assessing global disconnects. We find that of the SRLG pair combinations, 36 (70.6\%) SRLGs cause no local disconnects (besides the two due to $S_{M,L}$ and $S_{L,H}$) or global disconnects; 4 (7.8\%) cause additional local disconnects; while 14 SRLGs (25.5\%) are involved in global disconnects. This latter value suggests that global disconnects are not confined to a small number of critical SRLGs. 

\begin{table}[!t]
\caption{Double SRLG failure analysis results, incl. disconnects per trial and number of SRLGs causing disconnects.}
\label{tab:double-fails}
\centering
\begin{threeparttable}
\begin{tabular}{l|rr}
\toprule
Category & Count & Percent\\
\midrule
Trials (pair combinations) & 1275 & \\
No disconnects & 1164 & 91.3\% \\
Local disconnects & 96 & 7.5\% \\
Global disconnects & 15 & 1.2\% \\
\midrule
SRLG count & 51 & \\
SRLGs not involved in any disconnects\tnote{a} & 36 & 70.6\% \\
SRLGs involved in local disconnects\tnote{a} & 4 & 7.8\% \\
SRLGs involved in global disconnects & 14 & 25.5\% \\
\bottomrule
\end{tabular}
\begin{tablenotes}
\item[a] Does not consider pairs containing $S_{M,L}$ or $S_{L,H}$. \vspace{-16pt}
\end{tablenotes}
\end{threeparttable}
\end{table}

The cities and towns particularly affected by disconnects can also be investigated. \autoref{tab:node-city} summarises the multilayer presence (number of layers) and overall node degree in the holistic network for each MPoP. We also consider disconnect counts across the 1275 SRLG pairs. We see that state capitals are generally well protected, apart from Hobart. Most global failures affect regional cities, but Hobart is overrepresented. Tasmania's connectivity problems have been long documented, and such failures have been seen, \eg \cite{Hales_2022}. This justifies the need for investment in Tasmanian fibre infrastructure. 
More generally, this shows how this type of analysis can provide important insights into a country's network infrastructure.  

\begin{table}[!t]
\caption{Summary metrics for each MPoP in the multilayer model, and the number of local and global disconnects the MPoP experienced across 1275 trials of double SRLG failures.}
\label{tab:node-city}
\centering
\footnotesize
\begin{tabular}{l|rrrrrrrrrrrrr}
\toprule
MPoP & Node & Multilayer & Local & Global  \\
Location & Layers & Node Degree & Disconnects & Disconnects \\
\midrule
Adelaide & 6 & 37 & 0 & 0 \\
Brisbane & 6 & 37 & 0 & 0 \\
Canberra & 6 & 20 & 3 & 1 \\
Darwin & 6 & 20 & 4 & 2 \\
Hobart & 6 & 15 & \textbf{123} & \textbf{6} \\
Melbourne & 6 & 49 & 0 & 0 \\
Perth & 6 & 33 & 0 & 0 \\
Sydney & 6 & \textbf{57} & 0 & 0 \\
Mount Isa & 2 & 6 & 1 & 1 \\
Launceston & 2 & 6 & 4 & 3 \\
Port Hedland & 2 & 6 & 0 & 0 \\
Rockhampton & 2 & 8 & 1 & 1 \\
Townsville & 3 & 8 & 4 & 4 \\
\bottomrule
\end{tabular}
 \vspace{-9pt}
\end{table}

\begin{figure}[!t]
    \centering
    \includegraphics[width=0.44\textwidth]{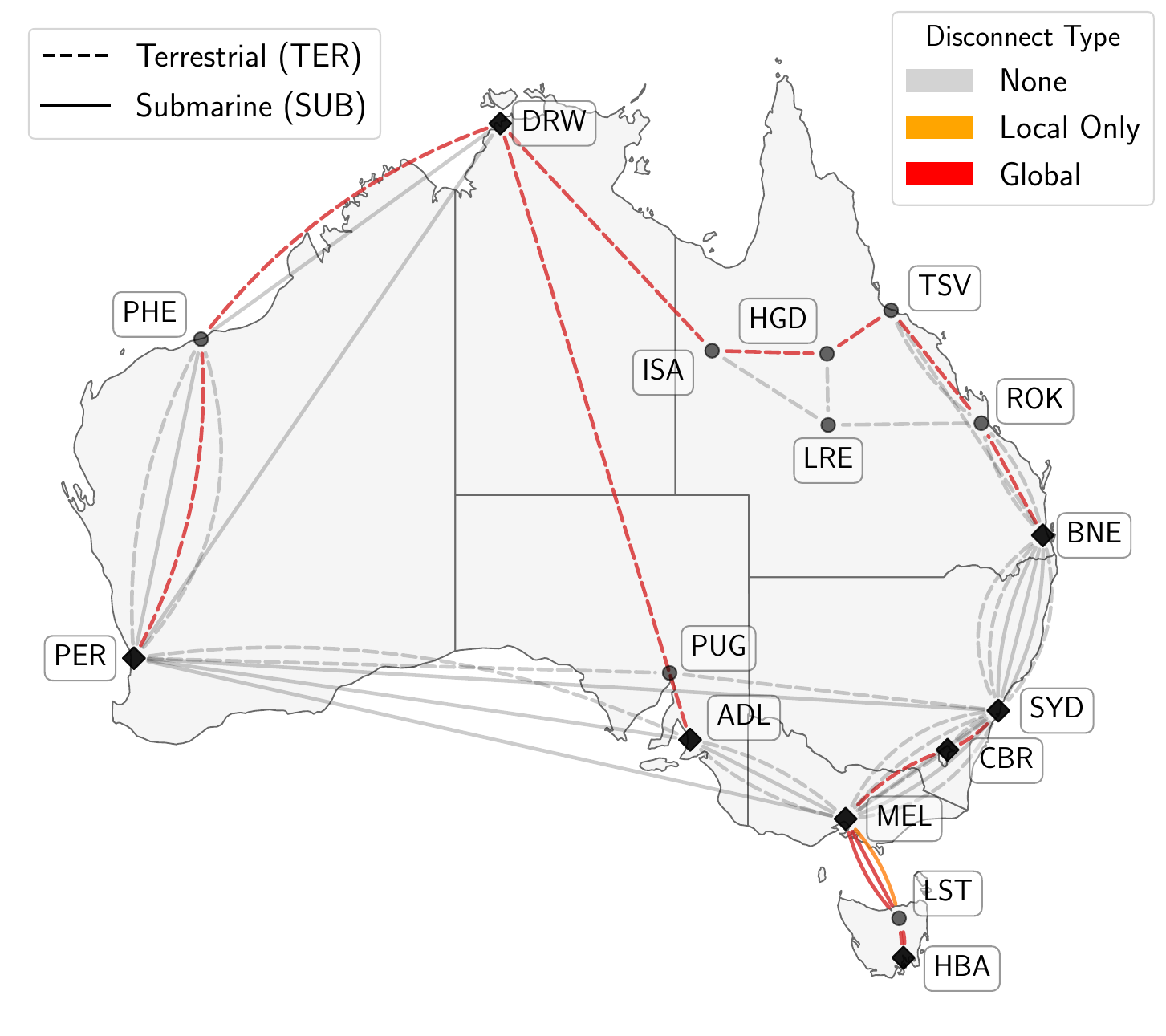}
    \caption{Locations of SRLGs involved in local and global disconnects during the double SRLG failure analysis.}
    \label{fig:fail-map}
    \vspace*{-5mm}
\end{figure}

Finally, we highlight the SRLGs affected by local and global disconnects on the Australian map shown in \autoref{fig:fail-map}. This again excludes the expected local disconnects caused by $S_{M,L}$ and $S_{L,H}$. We observe that Tasmanian SRLG removals clearly cause a considerable number of disconnects, both the submarine and terrestrial cables. Paths with fewer alternative SRLG routes, such as Adelaide to Darwin, are also impacted. This indicates specific locations requiring increased infrastructure.

\section{Summary and Future Work}
\label{sec:conclusion}
In this paper we study Australia's inter-city optical fibre network. We considered the impact of SRLGs on the robustness of Australia's Internet network, the effect of IXPs, and how diversity of ISP networks assists the network. We contribute a generalised multilayer network framework, an Australian-specific model, and a corresponding SRLG failure analysis of the Australian model. In future, we hope to verify ISP topology maps using traceroute surveys and quantify the likelihood of particular SRLG groupings. Investigating subsets of the network, such as performance in the absence of a specific layer, and the addition of new SRLGs that would provide the greatest benefit when implemented, are of interest. Finally, we wish to extend our failure analysis by incorporating Internet traffic flows over the network, to investigate the impact of rerouting policies and cascade failures.

\bibliographystyle{IEEEtran}  
\bibliography{ref}

\end{document}